%% file: paper_arxiv.tex
\documentclass{article}
\usepackage{arxiv}

\AtBeginDocument{%
  \providecommand\BibTeX{{%
    \normalfont B\kern-0.5em{\scshape i\kern-0.25em b}\kern-0.8em\TeX}}}

\usepackage{xspace}
\usepackage{algorithm}
\usepackage[noend]{algpseudocode}
\usepackage{subcaption}
\usepackage{makecell}
\usepackage{multirow}
\usepackage{url}

\usepackage{graphicx}

\title{Towards Fair Federated Recommendation Learning: Characterizing the Inter-Dependence of System and Data Heterogeneity}
\author{Kiwan Maeng \\
Meta AI \\
\And
Haiyu Lu \\
Meta \\
\And
Luca Melis \\
Meta \\
\And
John Nguyen \\
Meta \\
\And
Mike Rabbat \\
Meta AI \\
\And
Carole-Jean Wu \\
Meta AI}

\begin{document}


\maketitle




\newcommand{\sys}{RF$^2$\xspace}
\newcommand{\meta}{\emph{Company X}\xspace}
\newcommand{\notes}[2]{{\bf\textsf{\textcolor{#1}{#2}}}} 
\newcommand{\todo}[1]{\notes{red}{TODO: #1}}

\begin{abstract}
Federated learning (FL) is an effective mechanism for data privacy in recommender systems by running machine learning model training on-device. 
While prior FL optimizations tackled the data and system heterogeneity challenges faced by FL, they assume the two are independent of each other. This fundamental assumption is \emph{not} reflective of real-world, large-scale recommender systems --- \emph{data and system heterogeneity are tightly intertwined}. This paper takes a data-driven approach to show the inter-dependence of data and system heterogeneity in real-world data and quantifies its impact on the overall model quality and fairness. We design a framework, \sys, to model the inter-dependence and evaluate its impact on state-of-the-art model optimization techniques for federated recommendation tasks. We demonstrate that the impact on fairness can be severe under realistic heterogeneity scenarios, by up to 15.8--41$\times$ compared to a simple setup assumed in most (if not all) prior work. It means when realistic system-induced data heterogeneity is not properly modeled, the fairness impact of an optimization can be downplayed by up to 41$\times$.
The result shows that modeling realistic system-induced data heterogeneity is essential to achieving fair federated recommendation learning.
We plan to open-source \sys to enable future design and evaluation of FL innovations.   
\end{abstract}




\maketitle

\input{intro}
\input{bg}
\input{prod}
\input{bench}
\input{methodology}

\input{eval}
\input{related}
\input{conclusion}

\section*{Acknowledgement}
\input{acks}

\bibliographystyle{unsrt} 
\bibliography{bib}

\end{document}

%% file: intro.tex
\section{Introduction}



Recommender systems are a fundamental building block of modern internet services, empowering day-to-day applications. Recommender systems suggest videos on Netflix~\cite{netflix} and Youtube~\cite{youtube}, music on Spotify~\cite{spotify}, apps on the Google Play Store~\cite{widedeep}, and stories on Instagram~\cite{instagram}.
A recent study showed that 60\% of YouTube's and 75\% of Netflix's videos watched were selected based on recommender systems~\cite{mckinsey, underwood2019use, xie2018personalized}. They are one of the important machine learning workloads, comprising 50\% of the training~\cite{acun2020} and 80\% of inference cycles~\cite{arch_impl} at Meta in 2019.

While recommender systems were traditionally trained inside datacenters, recent studies are increasingly exploring training them on each client device, using \emph{federated learning} (FL)~\cite{fet, alibaba_fl}. 
FL is a privacy-enhancing training method that is already well-adopted in many commercial products for non-recommendation use-cases, including Google's Gboard~\cite{gboard_hard, gboard_yang} and Meta's Oculus keyboard~\cite{oculus_fl}.
FL trains a model locally on each client device using its local data and later aggregates only the model updates. FL does not require raw user data to leave the client device.  

Training models with FL faces several unique challenges due to the data and system heterogeneity of participating client devices.
\emph{Data heterogeneity} refers to the fact that data in each user device is not \emph{independent and identically distributed} (IID), which makes convergence more challenging~\cite{fl_open_problems}. \emph{System heterogeneity} refers to the fact that client devices (e.g., smartphones) have different system capabilities from each other, which generally limits the model capacity and training efficiency~\cite{fl_open_problems}. Both challenges have been extensively tackled by prior literature~\cite{fl_open_problems, heterofl, fjord, oort, expanding_reach, fl_comm, fl_kd, flash}.

However, no prior work looked at the \emph{inter-dependence} of data and system heterogeneity, assuming the two are independent of each other.
Prior work used a random mapping approach to model data and system heterogeneity simultaneously~\cite{heterofl, fjord, oort, flash}, which we show to always produce zero correlation between the two heterogeneity (Section~\ref{sec:prod:real-world}).
%
%
%
By analyzing data from a large-scale recommender system deployment, we show that the simplistic assumption is not representative of the real world -- in real systems, data and system heterogeneity are tightly intertwined (Section~\ref{sec:prod:real-world}).
We refer to the tight correlation as \textbf{\emph{system-induced data heterogeneity}}.
We show that the system-induced data heterogeneity in real data can cause optimizations to experience fairness issues, which is a phenomenon not observed in prior work.
To the best of our knowledge, this is the first time system-induced data heterogeneity and its effects are reported and demonstrated with real-world data.

Based on this important observation, we developed \sys (Realistic Federated Recommendation for Fairness), an FL framework for recommender systems that simulates various degrees of system-induced data heterogeneity. \sys includes: (1) code to simulate FL using popular recommendation models and datasets, (2) a statistical method to model and control system-induced data heterogeneity, and (3) implementations of various popular FL optimizations for system heterogeneity~\cite{fl_comm, heterofl, fjord, gboard_hard, fl_at_scale}.
Our evaluation with \sys reveals that several popular FL optimizations can hurt the model fairness severely in a realistic setup with system-induced data heterogeneity, sometimes by more than 40$\times$ compared to a simplistic case with no system-induced data heterogeneity.
Our evaluation also lists several interesting observations. For example, we show that methods that showed similar fairness implications with no system-induced data heterogeneity can show significantly different fairness impacts with realistic system-induced data heterogeneity. We also show optimizations that achieve the best accuracy are not always the fairest (e.g., two similar-accuracy optimizations can differ in their fairness by 4.88$\times$).
We hope our evaluation motivates the need to simulate more realistic system-induced data heterogeneity. \sys can aid future researchers in systematically studying the realistic fairness impact of their systems.
Our key contributions are:
\begin{enumerate}
    \item We identify the existence of system-induced data heterogeneity and its potential effects in real-world data. To the best of our knowledge, this work is the first to explicitly reveal such effects in the real world.
    \item We propose a method to synthesize system-induced data heterogeneity onto existing datasets. Datasets generated with our method can simulate interesting fairness effects of the real world, while prior approaches cannot.
    \item We present \sys, an FL simulation framework for recommendation models that can simulate system-induced data heterogeneity and various FL optimizations from prior work. \sys will be open-sourced upon publication, which will allow future researchers to assess their optimizations in a more realistic setup.
    \item Our evaluation lists several effects of system-induced data heterogeneity on existing optimizations. We hope the findings will inspire future researchers to design and evaluate fair FL systems on a more realistic setup.
\end{enumerate}



%% file: bg.tex
\section{Background and Motivation}

\subsection{Deep Learning Recommender Systems}

Recommender systems suggest items to users by predicting the likelihood of interaction (e.g., click or purchase) between a user and items.
We broadly use the term \emph{click} to refer to any positive user-item interaction. 
Various techniques have been explored to deliver high-quality recommendations, ranging from classical techniques, e.g., matrix factorization~\cite{mf}, to emerging new deep learning-based techniques~\cite{widedeep, din, dien, dlrm, dcn, dcnv2, deepfm}, just to name a few.
In this paper, we will focus on deep learning-based approaches and refer to them as recommender systems.

Deep learning-based recommender systems use features of users and items as inputs to predict whether a user will click a particular item.
Two commonly-used feature types are dense features and sparse features. Dense features represent features of continuous values, such as a user's age or the price of an item. Sparse features represent categorical features of discrete values, such as a user's gender, the collection of items a user liked in the past, or the genre of a movie.
Sparse features are usually encoded as an extremely sparse one- or multi-hot vector.

To predict the click probability, recommender systems first translate sparse features into dense embedding vectors using embedding tables~\cite{dlrm, din, widedeep}. The embedding vectors are merged with dense features and go through a multi-layer perception (MLP), producing a prediction at the end.
Different model architectures explore variations in how the features are merged, including simple concatenation~\cite{neumf}, element-wise multiplication~\cite{neumf}, pairwise dot product~\cite{dlrm, widedeep}, attention-based weighted averaging~\cite{din}, or using another deep model~\cite{dcnv2, dien}.

\subsection{Federated Learning}
\label{sec:bg:fl}

Federated learning (FL)~\cite{gboard_hard} trains a model using a pool of client devices without each client having to send its data to the server. In this section, we discuss the workflow of FL and how prior literature handles data and system heterogeneity.

\subsubsection{Workflow of Federated Learning}

To train a model using FL, a centralized server first selects clients to participate from the client pool. The selected clients download the model from the server and train it locally using their data.
After training, the clients upload their trained models (or equivalently, the gradients) back to the server. When all the participating clients upload their gradients, the server aggregates the gradients and updates the server-side model. The process repeats until the model converges.
In the most commonly-used FedAvg algorithm~\cite{gboard_hard}, the server aggregates client gradients using weighted averaging, where the number of samples in each client corresponds to a weight value. Then, the aggregated gradient is simply added to the server model or applied using a separate server-side optimizer~\cite{fieldguide}.
%

\subsubsection{Data Heterogeneity}
\label{sec:bg:data_hetero}

FL is a form of distributed ML training. However, unlike distributed training in datacenters where data can be shuffled so that each trainer node has an independent and identically distributed (IID) subsample~\cite{ps}, the data of each FL client is non-IID --- the number of samples and the feature/label distributions on each client are different from each other~\cite{fl_open_problems}.
Data heterogeneity makes it challenging to reach high model quality~\cite{dirichlet}. Many algorithms~\cite{gboard_hard, fednova, fedprox} have been proposed to improve the model quality in the presence of data heterogeneity.

\begin{figure}
     \centering
     \begin{subfigure}[b]{0.42\textwidth}
         \centering
         \includegraphics[width=\textwidth]{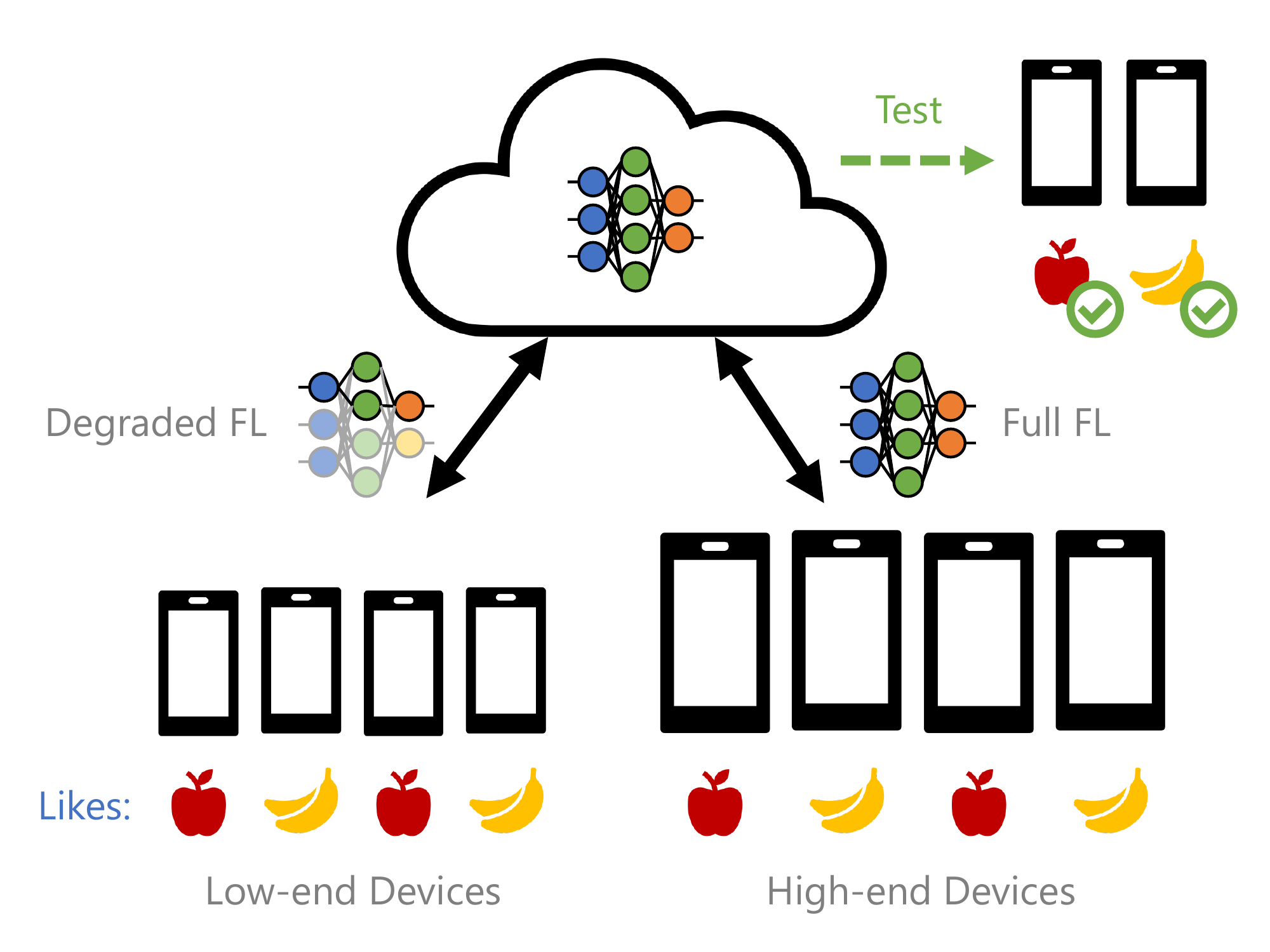}
         \caption{A case without system-induced data heterogeneity}
         \label{fig:motivation0}
     \end{subfigure}
     \begin{subfigure}[b]{0.42\textwidth}
         \centering
         \includegraphics[width=\textwidth]{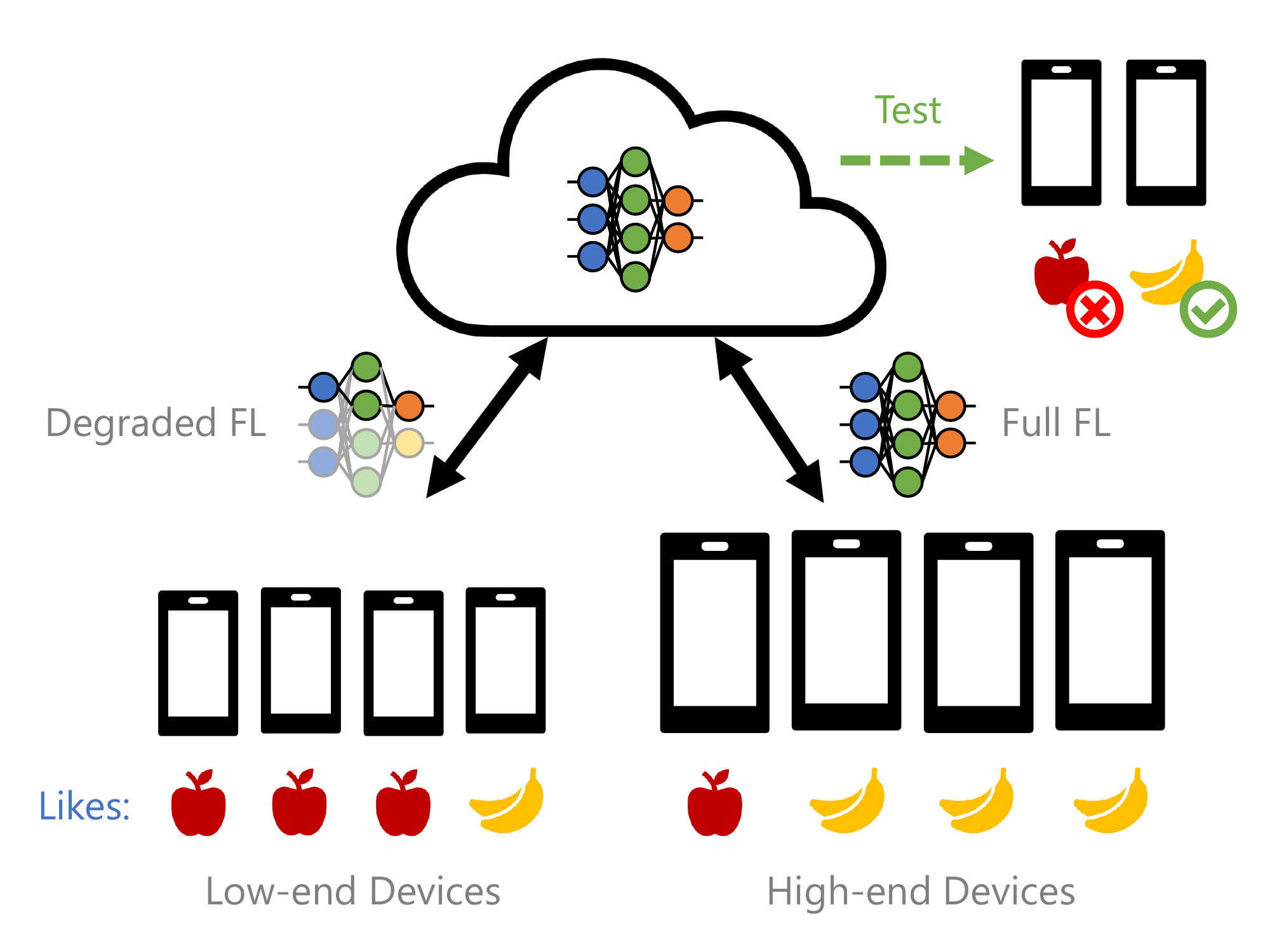}
         \caption{A case with system-induced data heterogeneity}
         \label{fig:motivation1}
     \end{subfigure}
        \caption{Tier-aware optimizations can hurt fairness when system-induced data heterogeneity is present. The figure shows an optimization that makes low-end devices train only a subset of the model~\cite{heterofl, fjord, expanding_reach}. The optimization produces a fair model if the data distribution between low- and high-end devices are similar (a), but may become unfair if the data distribution differ significantly (b).
        }
        \label{fig:motivation}
\end{figure}

\subsubsection{System Heterogeneity and Tier-Aware Optimizations}
\label{sec:bg:system_hetero}

Client devices (e.g., smartphones) vary significantly in their system capabilities, including computing power, memory, storage, and network speed~\cite{fedscale, flash, system-heterogeneity}. For example, low-end and high-end smartphones may experience a 2--6$\times$ latency difference when training the same model~\cite{flash} and two orders of magnitude difference in their network bandwidth~\cite{speedtest}.
The \emph{system heterogeneity} degrades the efficiency of FL because each round in FL proceeds only after all the participating clients finish training and send back updates.
The synchronous nature makes slow clients become \emph{stragglers} that bottleneck the entire training process.

To mitigate the straggler effect, recent studies proposed \emph{tier-aware optimizations}.
The core idea is to group devices with similar system capabilities into \emph{tiers} and apply distinct optimizations to different device tiers, so that lower-tier devices bear lighter computation/communication burdens. Below, we describe some of the commonly proposed forms:

\paragraph{\bf Excluding low-end devices}
The simplest optimization is to prevent low-end devices from participating in FL entirely to minimize the presence of stragglers. This simple solution can either be implemented by explicitly leaving out low-end devices~\cite{gboard_hard} or by implicitly setting a training time deadline that low-end devices cannot meet~\cite{fl_at_scale}. Many real products have adopted this strategy. For example, Google's Gboard's next-word prediction disallows devices with less than 2GB RAM from participating in their FL process~\cite{gboard_hard}).

\paragraph{\bf Over-selection and dropping}

Another well-adopted optimization is to select $N$\% more clients than needed during selection and drop the slowest $N$\% during aggregation~\cite{fedbuff}. Low-end devices are more likely to be dropped by this optimization because they are more likely to end up being the slowest $N$\%.

\paragraph{\bf Tiered gradient compression}

When there is a network bandwidth imbalance between tiers, applying gradient compression (e.g., gradient pruning~\cite{fl_comm, adaptive_fl_dropout, adaptive_fl_pruning, hetero_pruning1} or quantization~\cite{fl_comm, qsgd}) more aggressively to devices with a slower network can balance the communication speed.
Not all techniques from other use-cases are applicable to FL, however. For example, the popular Top-K pruning~\cite{topk} may leak which entries of the embedding tables were accessed in FL~\cite{alibaba_fl}.

\paragraph{\bf Tiered model sizes}

When model computation time imbalance is severe, using smaller models for devices with less computing capabilities can relieve the imbalance.
Several prior work proposed using a smaller number of channels for low-tier devices to reduce computation time and memory usage~\cite{heterofl, fjord, expanding_reach}.
Upon model aggregation, channels are only averaged across tiers that use the channels~\cite{heterofl, fjord, expanding_reach}, and knowledge distillation can be additionally used to further improve model accuracy~\cite{fjord}.
Others allowed each device tier to use an entirely different model from each other and relied on knowledge distillation to aggregate the knowledge~\cite{feddf, fjord, fedgkt, cronus, fedmd, faug, yaejee}.
Figure~\ref{fig:motivation0} illustrates an example of a tier-aware optimization, where low-end devices train a smaller model with fewer channels in its hidden layers.

%% file: prod.tex
\section{Real-world Observations: Data and System Heterogeneity are Intertwined}
\label{sec:prod}

\begin{figure}
     \centering
     \begin{subfigure}[b]{0.33\textwidth}
         \centering
         \includegraphics[width=\textwidth]{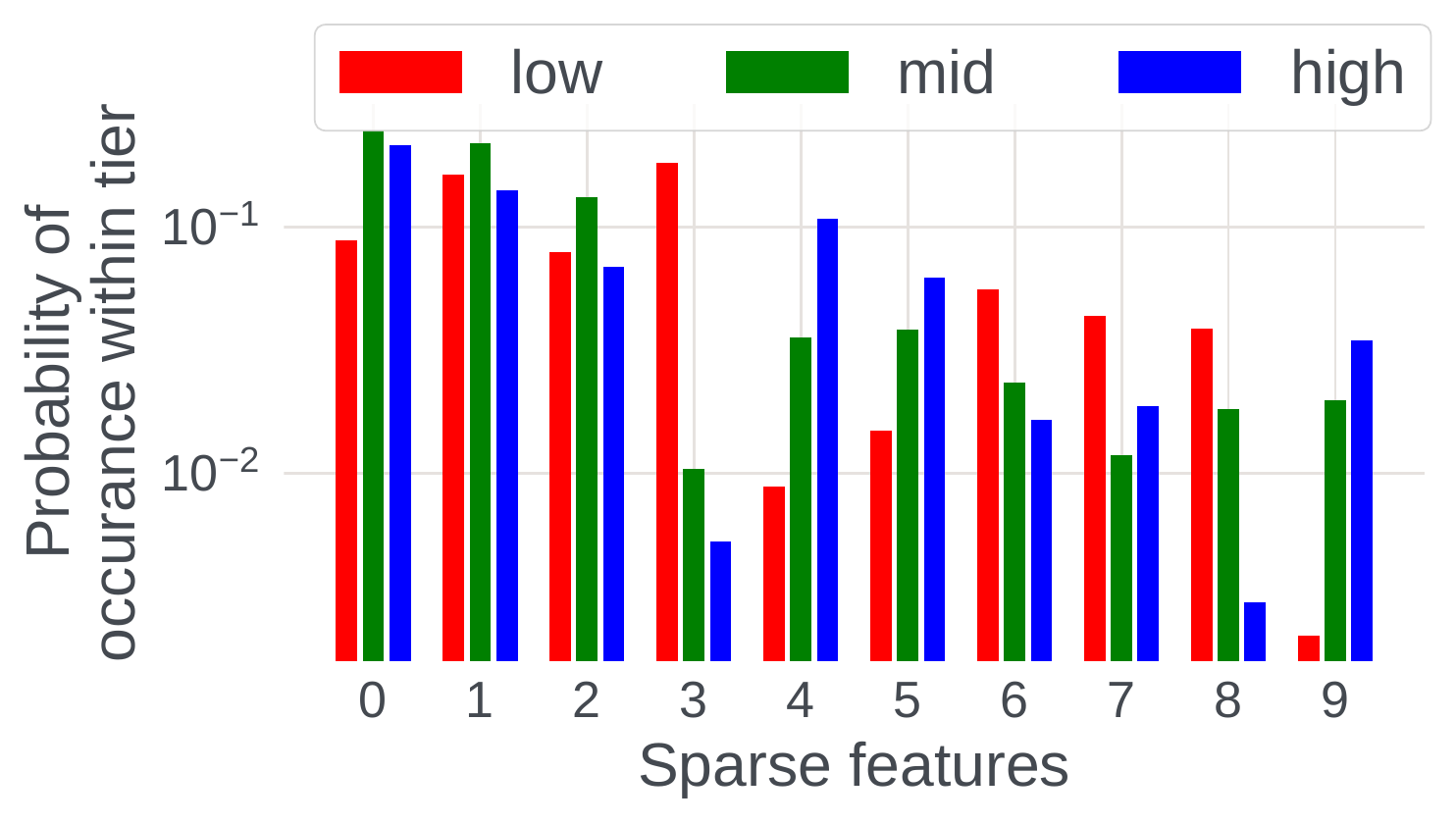}
         \caption{Real-world statistics}
         \label{fig:prod_0}
     \end{subfigure}
     \begin{subfigure}[b]{0.33\textwidth}
         \centering
         \includegraphics[width=\textwidth]{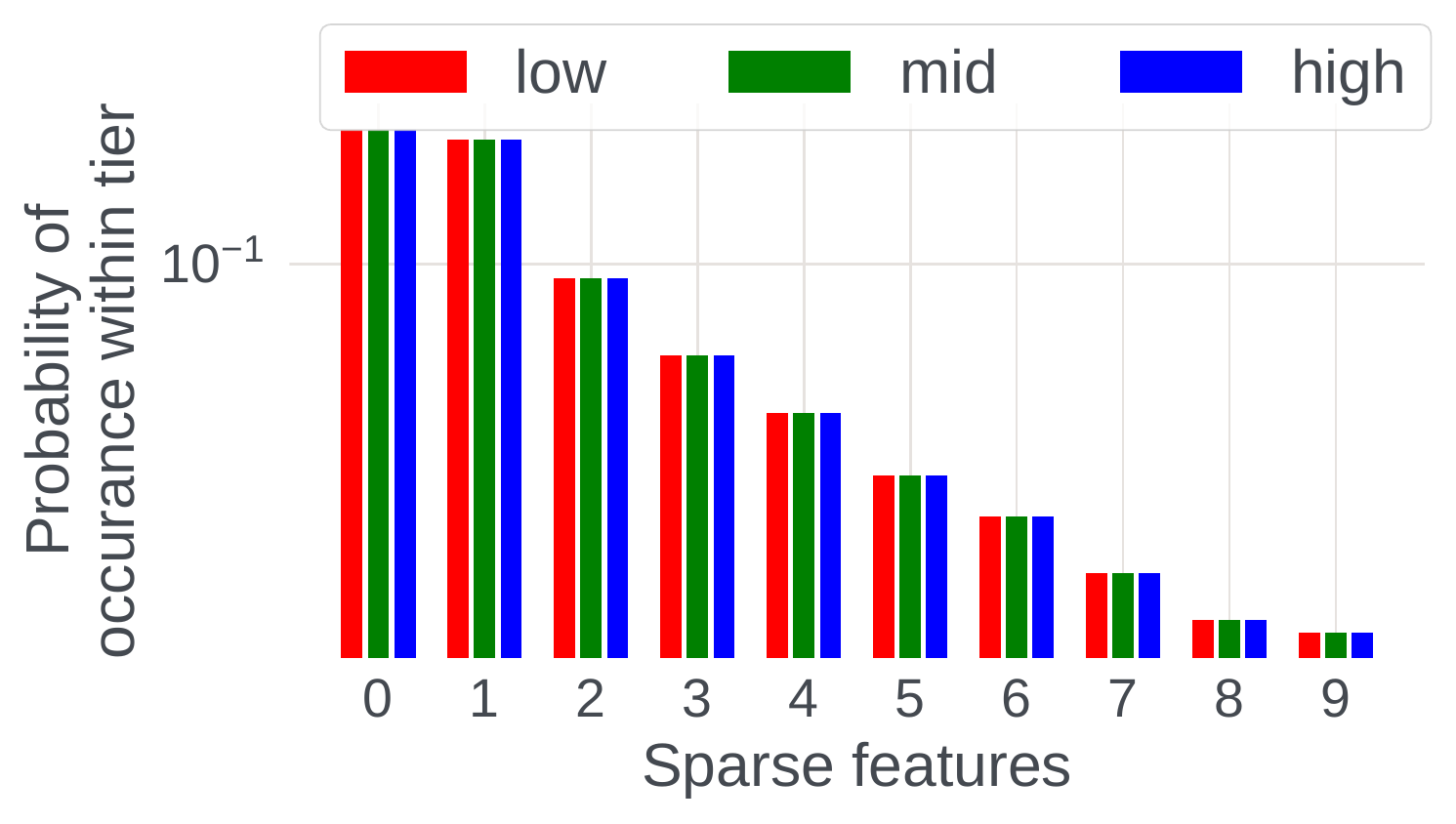}
         \caption{Random-mapping statistics}
         \label{fig:prod_1}
     \end{subfigure}
     \begin{subfigure}[b]{0.33\textwidth}
         \centering
         \includegraphics[width=\textwidth]{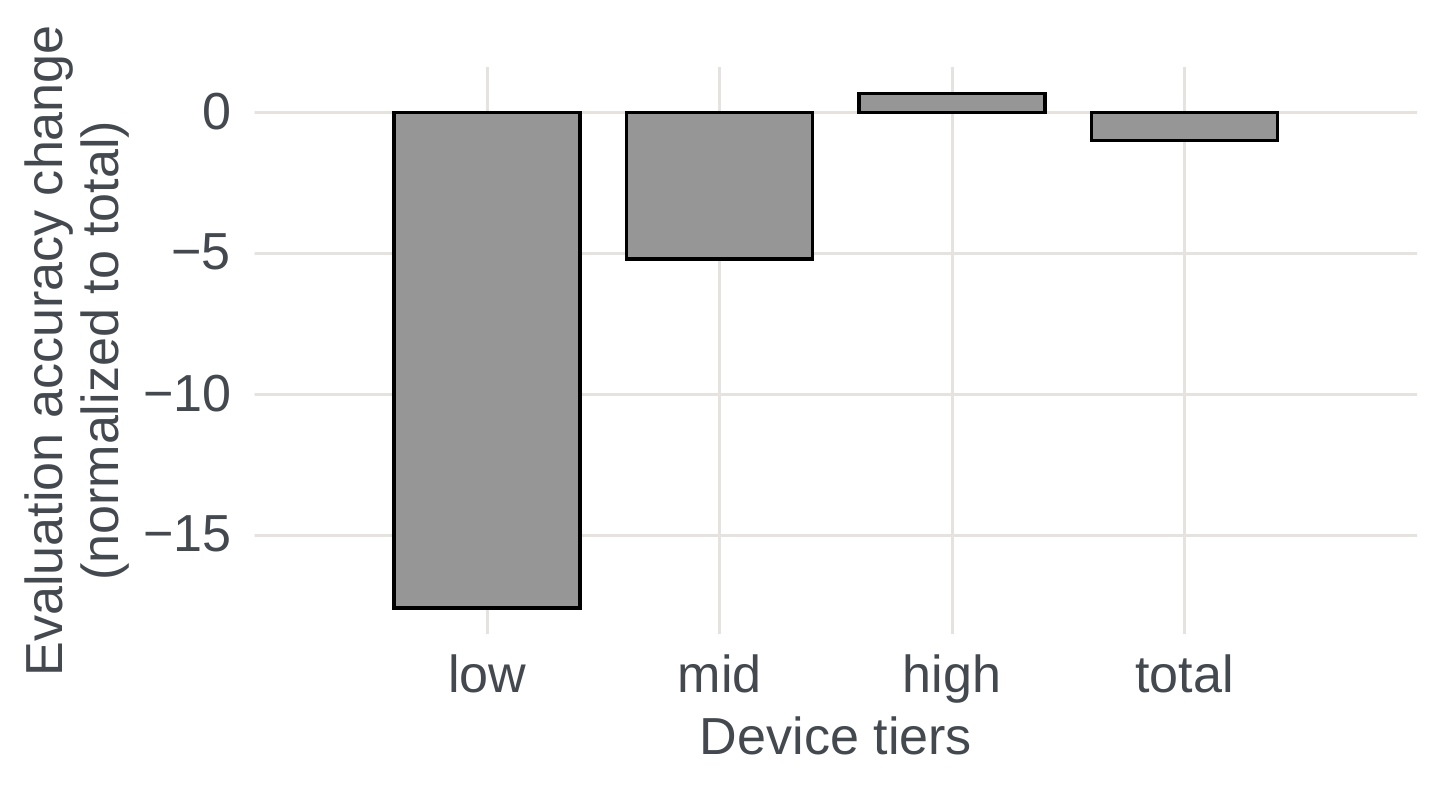}
         \caption{Fairness implications}
         \label{fig:prod_unfair}
     \end{subfigure}
        \caption{Real world experiences system induced data heterogeneity that can impact fairness. (a) plots the distribution of feature values across tiers in the real world, and (b) plots what the distribution would look like instead with random tier mapping. Comparing the two clearly shows that the real world experiences system-induced data heterogeneity.
        (c) shows the accuracy change for each tier when excluding low-end devices from training, in the presence of system-induced data heterogeneity in (a). Low-end devices are disproportionately penalized compared to the overall population.
        }
        \label{fig:prod}
\end{figure}

\subsection{Inter-dependence Between Data and System Heterogeneity}

Section~\ref{sec:bg:data_hetero}--\ref{sec:bg:system_hetero} discussed a stream of prior research that tackled the data heterogeneity and system heterogeneity of FL.
However, most (if not all) prior studies tackled data and system heterogeneity separately, assuming \emph{no inter-dependence exists between the two}. This assumption, however, is not reflective of the real world.

As a motivating example, assume that there are clients who like apples and clients who like bananas in the world, and their fruit preferences are an important feature of a recommender system (e.g., the system recommends apple juice to apple-liking clients). If the probability of liking apples or bananas is the same regardless of the client's device tier as in Figure~\ref{fig:motivation0}, we say there exists no inter-dependence between data and system heterogeneity.

Alternatively, there can be cases where the probability of liking apples is higher for low-end devices, while the probability of liking bananas is higher for high-end devices (Figure~\ref{fig:motivation1}). When there is such data distribution difference \emph{between device tiers}, inter-dependence exists between data and system heterogeneity. We term such an inter-dependence as \textbf{\emph{system-induced data heterogeneity}} in this work.
When system-induced data heterogeneity exists, applying tier-aware optimizations may cause fairness issues. For example, if we use fewer channels for low-end devices, and low-end devices mostly hold apple-liking features, the final trained model may not work as well for apple-liking clients as for banana-liking clients, because most of the apple-liking data were trained through a model with fewer channels (Figure~\ref{fig:motivation1}).
We show in Section~\ref{sec:prod:real-world} that real-world recommender systems experience system-induced data heterogeneity, and the fairness of the model can be impacted when tier-aware optimizations are applied without careful considerations. Thus, it is important to simulate realistic system-induced data heterogeneity when evaluating FL optimizations.

Unfortunately, no prior FL literature assumed the existence of system-induced data heterogeneity to the best of our knowledge, and hence no prior optimizations were evaluated in the presence of realistic system-induced data heterogeneity. When simulating data and system heterogeneity, even the advanced FL simulators with real-world system traces~\cite{fedscale, flash}, synthesized or collected client data that show data heterogeneity and \emph{randomly assigned} synthesized or collected system heterogeneity characteristics (i.e., device tiers) to each client~\cite{fjord, heterofl, fedscale, flash}.
Such a random tier-client mapping always produces a dataset with no system-induced data heterogeneity, as in Figure~\ref{fig:motivation0}.
%

\subsection{Does the Real World Experience System-induced Data Heterogeneity?}
\label{sec:prod:real-world}

To understand whether system-induced data heterogeneity exists in the real world, we analyzed important sparse features of a recommender system that serves billions of users worldwide.
Figure~\ref{fig:prod_0} presents the statistics of a sparse feature that is known to be important in delivering high-quality recommendations (e.g., fruit preferences in Figure~\ref{fig:motivation}). 
We group the user devices into three tiers (low-, mid-, and high-) based on their system capabilities and observe how frequently each value in the feature (e.g., apple, banana, orange, ...) occurs within each tier.
Figure~\ref{fig:prod_0} plots the result for the top-10 most frequently observed values. In Figure~\ref{fig:prod_1}, we plot the statistics again, but this time, by mapping users to tiers randomly as was done by the prior work~\cite{heterofl, fjord, oort, fedscale, flash} instead of using the actual tiers.

Comparing Figure~\ref{fig:prod_0} and Figure~\ref{fig:prod_1}, it is clear that \textit{real-world deployment environment experiences notable system-induced data heterogeneity}.
When using random tier mapping (Figure~\ref{fig:prod_1}), the probability of each sparse feature value occurring is the same across tiers. In other words, the affinity to apples/bananas is the same across device tiers (Figure~\ref{fig:motivation0}).
However, real-world data (Figure~\ref{fig:prod_0}) exhibits high data heterogeneity across tiers, closely resembling the scenario depicted in Figure~\ref{fig:motivation1}. For example, sparse feature value 3 is mostly observed only in the low-end device tier, resembling the preferences for apple in Figure~\ref{fig:motivation1}. Value 9, on the other hand, is mostly observed in the mid/high-end device tiers but very scarcely in the low-end device tier, resembling the preferences for bananas in Figure~\ref{fig:motivation1}.

We also demonstrate with that popular tier-aware optimizations can introduce fairness degradation in the presence of system-induced data heterogeneity. We trained a recommendation model using data similar to Figure~\ref{fig:prod_0} while prohibiting low-end devices from participating, following Google's Gboard FL setup~\cite{gboard_hard}.
Figure~\ref{fig:prod_unfair} shows the resulting model accuracy change for each tier, normalized by the overall average.
Low-end devices get disproportionately affected by the optimization, suffering from 17.6$\times$ more accuracy degradation than the average population.
%
Figure~\ref{fig:prod_unfair} motivates the need to study tier-aware optimizations in a realistic system-induced data heterogeneity setup. If not, model prediction quality for certain populations can be significantly degraded unintentionally.


%% file: bench.tex
\section{Studying System-induced Data Heterogeneity for Recommender Systems}

\begin{figure*}
    \centering
    \includegraphics[width=0.95\textwidth]{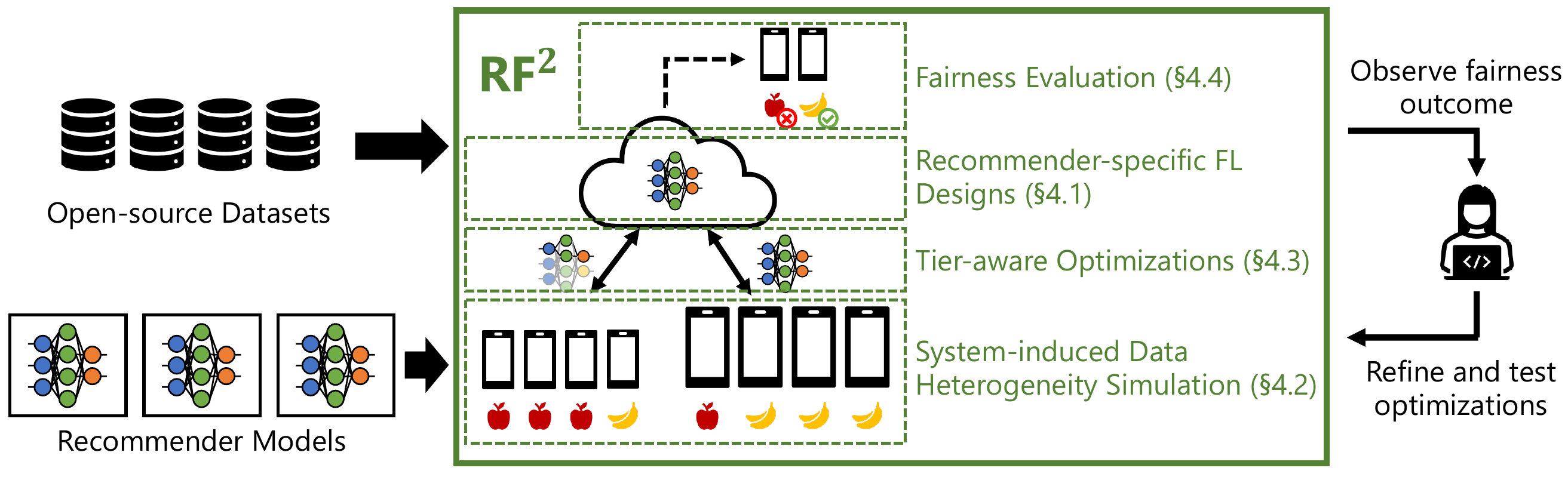}
    \caption{Overview of \sys.}
    \label{fig:sys}
\end{figure*}

\textbf{\sys} is an FL simulation framework for recommender models that enables agile modeling of system-induced data heterogeneity. \sys supports (1) efficient FL training for popular recommender models and datasets (Section~\ref{sec:sys:models}), (2) synthesizing varying degrees of system-induced data heterogeneity onto existing datasets (Section~\ref{sec:sys:sidh}), (3) a family of tier-aware optimization strategies from prior work (Section~\ref{sec:opt}), and (4) fairness evaluation (Section~\ref{sec:bench_fairness}) that can guide programmers to refine and test their optimizations. 
Figure~\ref{fig:sys} illustrates the design overview of \sys. 

\subsection{Simulating FL for Recommender Systems}
\label{sec:sys:models}

\sys supports FL simulation for state-of-the-art, commonly-used recommender models and datasets.
While FL for deep recommender models has been studied in previous literature~\cite{alibaba_fl, fedncf}, prior frameworks were either confined to simplistic models that take only user ID and item ID as inputs~\cite{fedncf} or were built on proprietary datasets~\cite{alibaba_fl}.
\sys, on the other hand, is compatible with a large body of popular recommender models, by being built on top of DeepCTR-Torch~\cite{deepctr-torch}, an open-source codebase that implements 19 recommender models (in a non-FL context) and is easily extensible to more.
\sys also currently supports two commonly-used open-source datasets, Taobao Ad Display/Click Data~\cite{taobao} and MovieLens-20M~\cite{movielens} (Section~\ref{sec:setup:dataset-models}), and can be extended to support additional datasets.

\sys makes some unique design decisions to improve convergence and model a more realistic setup.
Instead of using minibatch SGD on the client~\cite{alibaba_fl, fedncf}, \sys implements an option to use a full-batch SGD.
Full-batch SGD is practical because recommender systems tolerate a large batch size,~\footnote{https://github.com/mlcommons/training\_results\_v1.1/blob/main/NVIDIA/benchmarks/dlrm/implementations/hugectr uses a batch size of 70k} and clients usually do not have many datapoints as user-item interaction is rare. For example, the Taobao dataset~\cite{taobao} has only 26 datapoints on average per client.
Using full-batch SGD on the clients and advanced optimizers, e.g., AdaGrad, on the server~\cite{fieldguide} improves the learning stability significantly.
%
%
\sys does not select a client again before every client is selected exactly once, unlike prior work that models duplicated selection~\cite{alibaba_fl, leaf, fedncf}.
The non-duplicate selection is to simulate a more realistic large-scale FL, where billions of clients participate~\cite{papaya, fl_at_scale} and duplicated selection is extremely rare.
%

\subsection{Simulating System-aware Data Heterogeneity}
\label{sec:sys:sidh}

One of \sys's main goals is to simulate realistic system-induced data heterogeneity. There are many potentially viable ways to simulate system-induced data heterogeneity. Across tiers, one can vary the distribution of user features, click rates, number of samples, or affinity to different items.
We concentrate on making the \emph{affinity to different items} heterogeneous across tiers (e.g., make certain tiers like certain items more, as in Figure~\ref{fig:motivation1}).
Our approach is applicable to any recommender datasets as they always have click information that represents the user-item affinity.
%

\begin{algorithm}
\begin{algorithmic}[1]
\State {$seenUsers, tier_{0}, tier_1, tier_2 \gets \emptyset, \emptyset, \emptyset, \emptyset$}
\For {$item \in items.sortedByDescendingPopularity()$} 
    \State {$p_0, p_1, p_2 \gets Dirichlet(\alpha, 3)$}
    \State {$p_{large}, p_{mid}, p_{small} \gets sortDescending(p_0, p_1, p_2)$}
    \State {$tier_{large}, tier_{mid}, tier_{small} \gets sortByDescendingSize(tier_0, tier_1, tier_2)$}
    \For {$user \in item.clickedUsers()$}
        \If {$user \notin seenUsers$}
            \State {$r \gets random()$}
            \State \algorithmicif\ {$r < p_{large}$} \algorithmicthen\ {$tier_{small} \gets tier_{small} \cup user$}
            \State \algorithmicelse\ \algorithmicif\ {$p_{large} \le r < p_{large} + p_{mid}$}
                \algorithmicthen\ {$tier_{mid} \gets tier_{mid} \cup user$}
            \State \algorithmicelse\
                \algorithmicthen\ {$tier_{large} \gets tier_{large} \cup user$}
            \State {$seenUsers \gets seenUsers \cup user$}
        \EndIf
    \EndFor
\EndFor
\end{algorithmic}
\caption{Dirichlet-based tier mapping.}
\label{alg:dirichlet}
\end{algorithm}

Algorithm~\ref{alg:dirichlet} shows how we assign tiers to each client to simulate system-induced data heterogeneity. Here, we assume three tier groups, $tier_0$, $tier_1$, and $tier_2$. Starting from the most popular item (Line 2), we draw three samples for the three tiers from a Dirichlet distribution~\cite{dirichlet} with a given $\alpha$ (Line 3).
$p_0$, $p_1$, and $p_2$ represent the probability for each user who clicked this item to be in each tier. If $\alpha$ is small, the values are more skewed, leading to higher system-induced data heterogeneity. If $\alpha$ is high, system-induced data heterogeneity is reduced.
We sort the three probabilities (Line 4) and also the number of already assigned users for each tier (Line 5), so that the tier with currently the least users gets the highest probability of the user being assigned. Lines 4--5 ensure that the final number of users is similar across tiers, and can be omitted if balancing the number of users is undesired.
With the given probability, each user that clicked the item (Line 6) gets assigned to one of the three tiers (Line 9--11), unless it is already assigned to a certain tier (Line 7). For users that never clicked any items, we treat them as clicking a null item and apply the same procedure.
Our tier assignment procedure is inspired by the approach used to simulate data heterogeneity in FL across clients~\cite{dirichlet}. Our algorithm has a different goal, which is to simulate system-induced data heterogeneity (data heterogeneity \emph{across tiers}).

\begin{figure}
     \centering
     \begin{subfigure}[b]{0.33\textwidth}
         \centering
         \includegraphics[width=\textwidth]{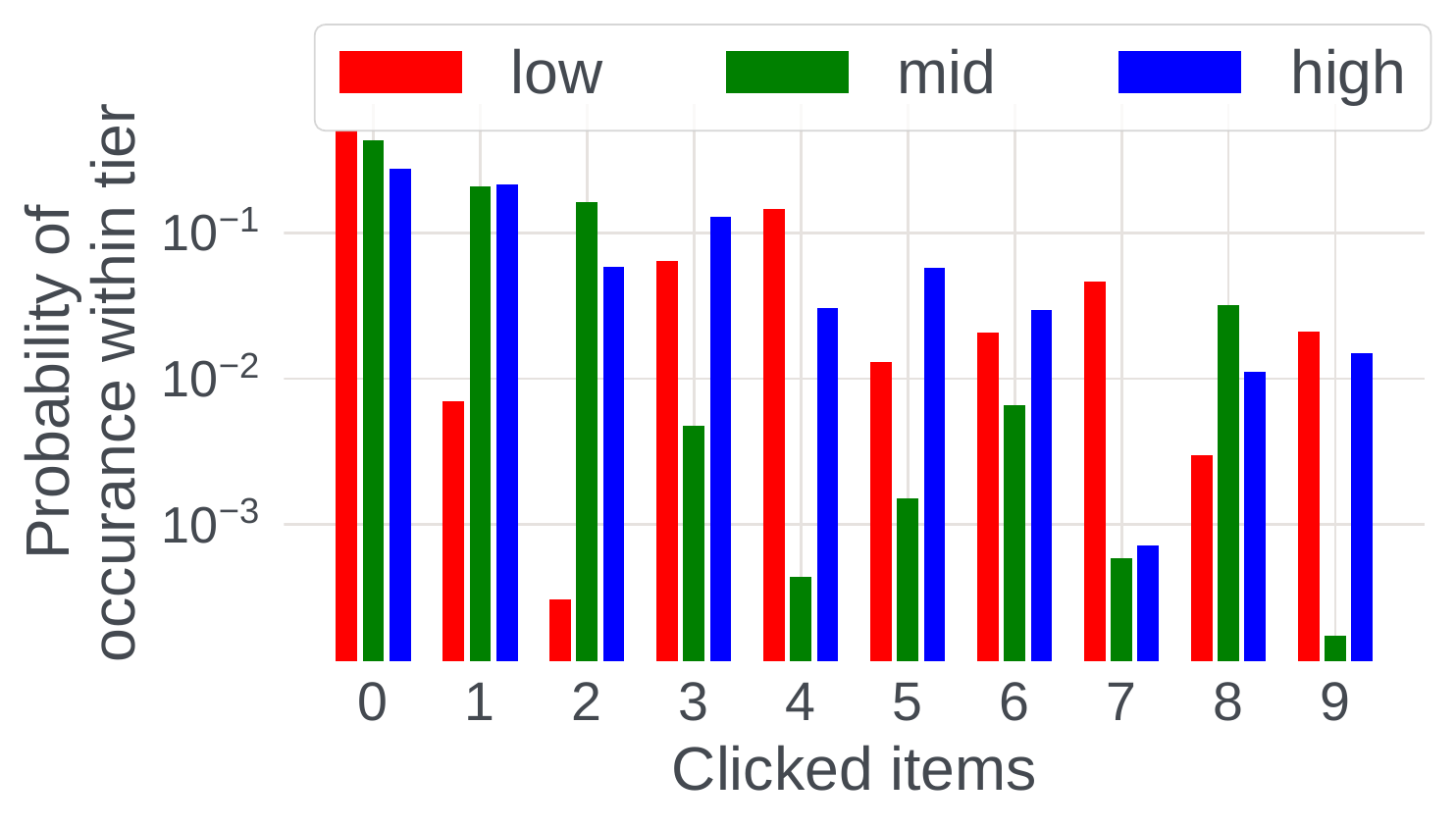}
         \caption{$\alpha=0.5$}
         \label{fig:dirichlet_0}
     \end{subfigure}
     \begin{subfigure}[b]{0.33\textwidth}
         \centering
         \includegraphics[width=\textwidth]{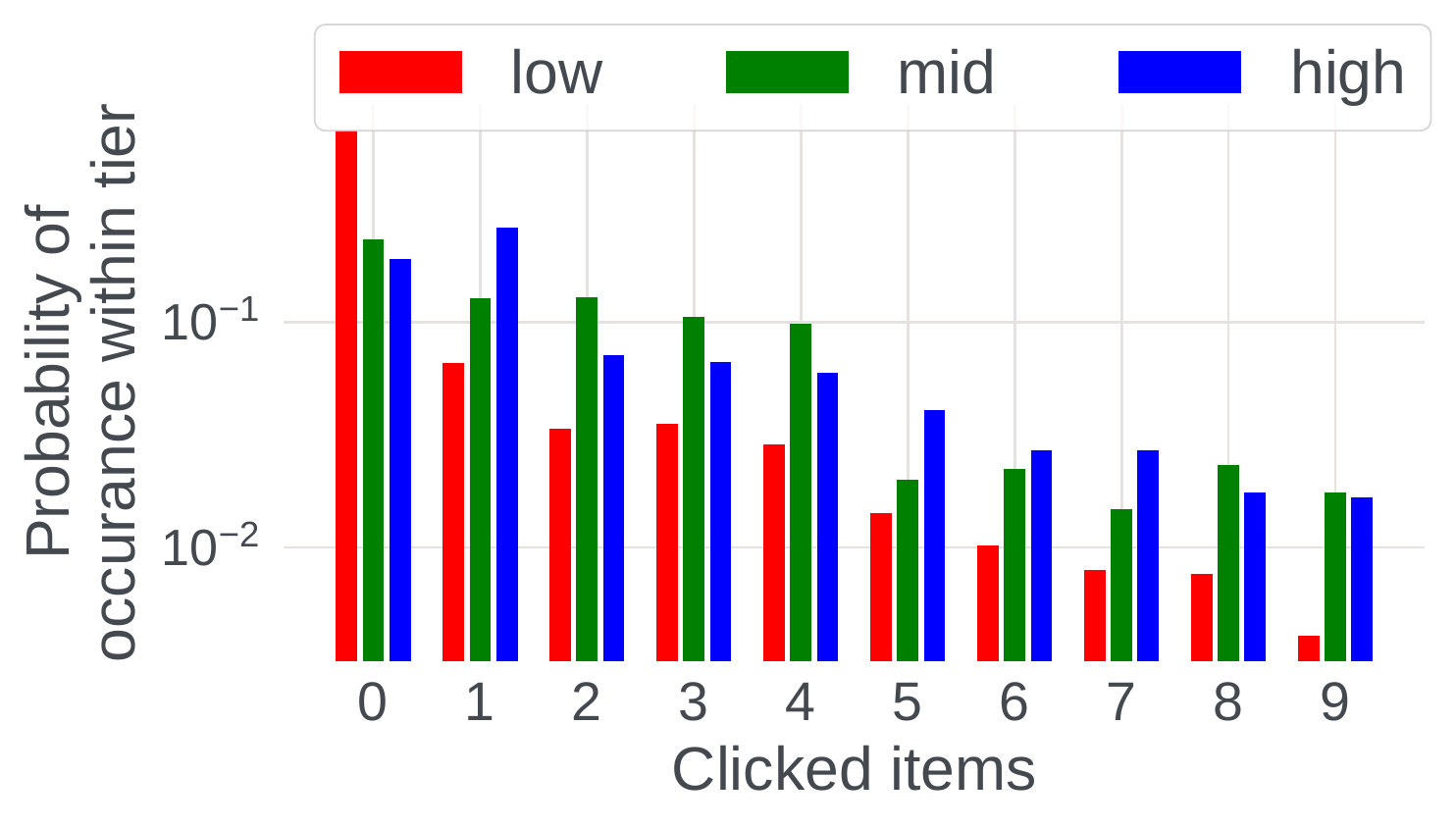}
         \caption{$\alpha=5$}
         \label{fig:dirichlet_1}
     \end{subfigure}
     \begin{subfigure}[b]{0.33\textwidth}
         \centering
         \includegraphics[width=\textwidth]{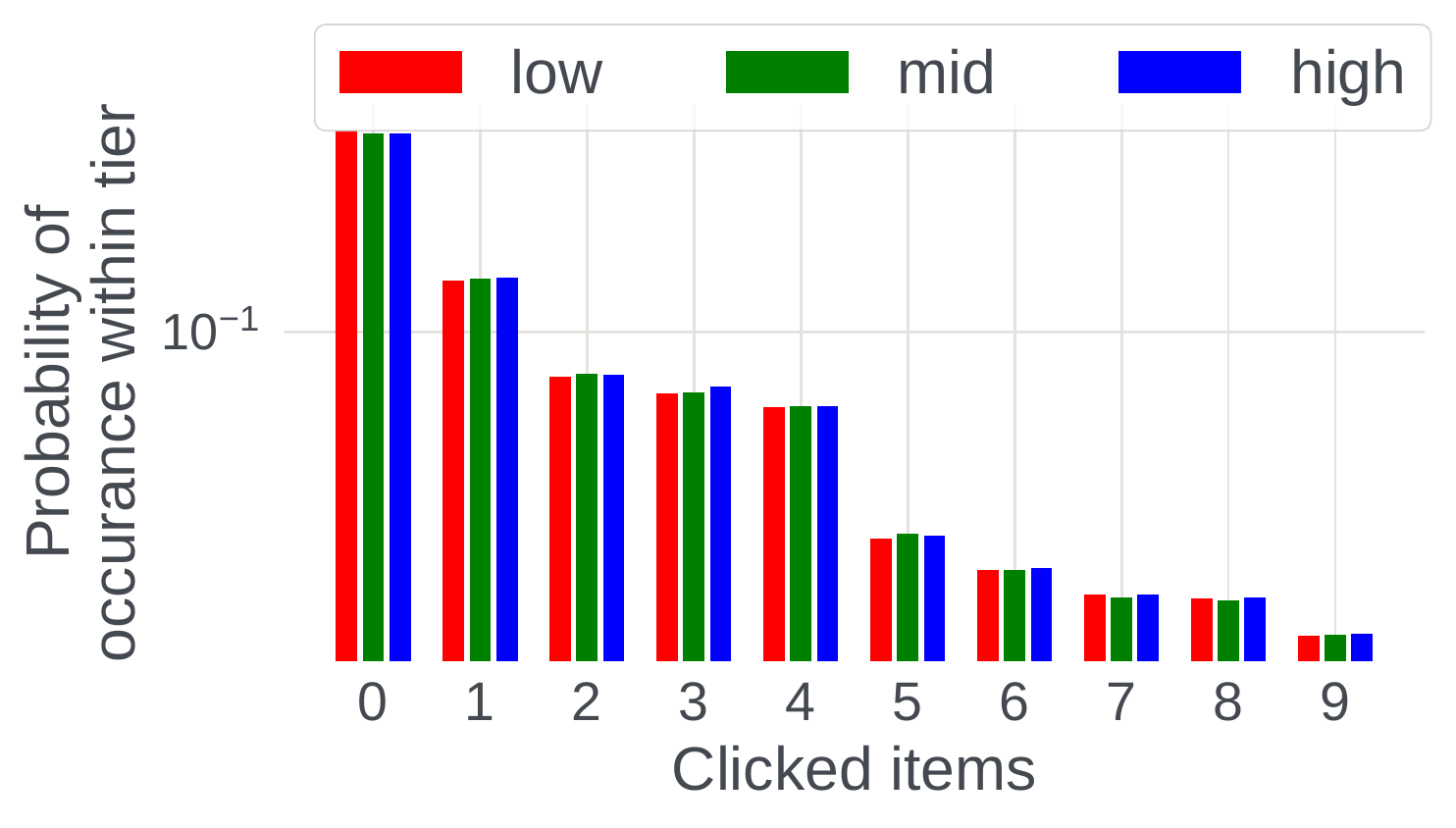}
         \caption{$\alpha=5000$}
         \label{fig:dirichlet_2}
     \end{subfigure}
        \caption{Dirichlet distribution with different $\alpha$ can simulate different degrees of system-induced data heterogeneity. 
        }
        \label{fig:dirichlet}
\end{figure}

\begin{figure}
     \centering
     \begin{subfigure}[b]{0.33\textwidth}
         \centering
         \includegraphics[width=\textwidth]{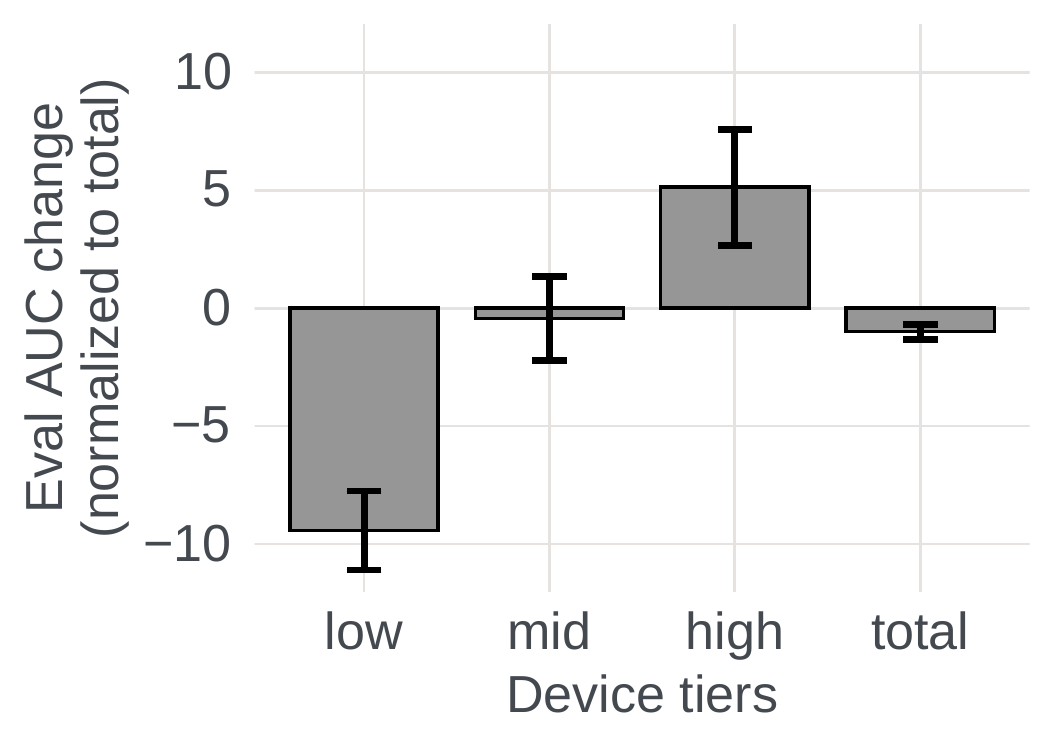}
         \caption{$\alpha=0.5$}
         \label{fig:taobao_unfair0_5}
     \end{subfigure}
     \begin{subfigure}[b]{0.33\textwidth}
         \centering
         \includegraphics[width=\textwidth]{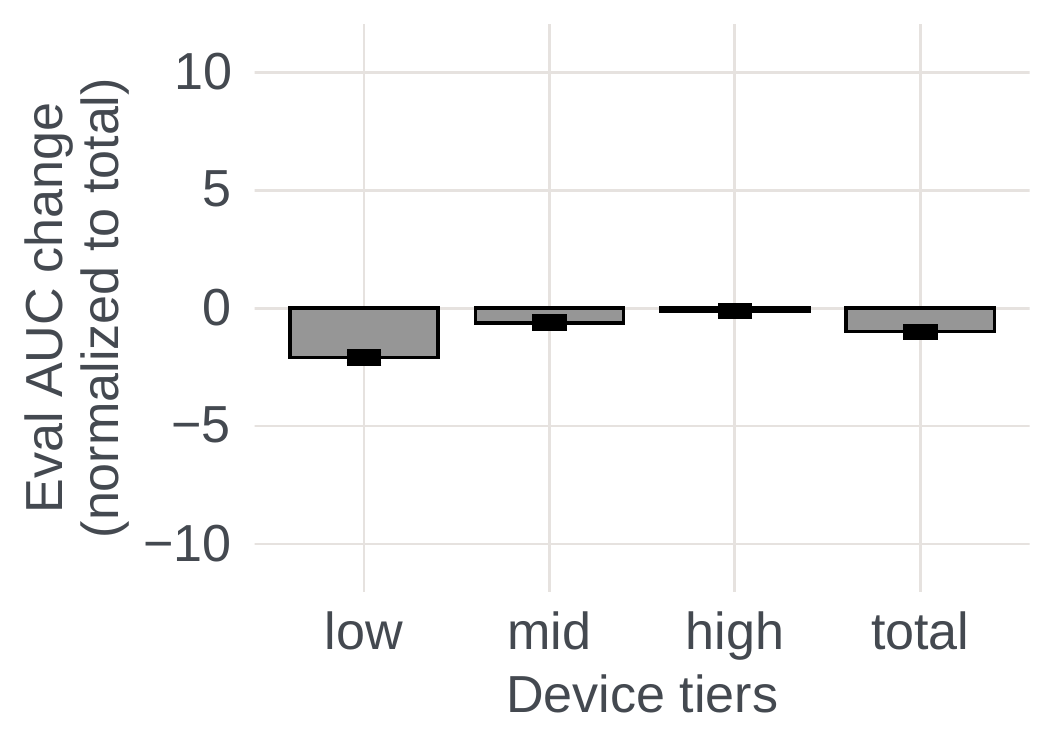}
         \caption{$\alpha=5000$}
         \label{fig:taobao_unfair5000}
     \end{subfigure}
        \caption{Tier-aware optimizations have a higher fairness impact when there is higher system-induced data heterogeneity, if we exclude low-end devices from training for the Taobao dataset.}
        \label{fig:taobao_unfair}
\end{figure}

Figure~\ref{fig:dirichlet} shows the generated system-induced data heterogeneity using different $\alpha$ for the MovieLens-20M dataset~\cite{movielens} (see Section~\ref{sec:setup:dataset-models} for more details on the dataset).
In the figure, the top-10 most clicked items and their occurrence on each (synthesized) tier are plotted.
We can see that for low $\alpha$ (Figure~\ref{fig:dirichlet_0}), the dataset experiences a severe system-induced data heterogeneity similar to that of the real world (Figure~\ref{fig:prod_0}). As we increase  $\alpha$ (Figure~\ref{fig:dirichlet_1}--~\ref{fig:dirichlet_2}), the distribution becomes increasingly more similar to random mapping (Figure~\ref{fig:prod_1}). 
The simulated system-induced data heterogeneity also leads to different fairness implications.
Figure~\ref{fig:taobao_unfair} shows an accuracy degradation of each tier when excluding low-end devices from FL under different system-induced data heterogeneity (see Sections~\ref{sec:methodology} and~\ref{sec:eval} for details on the evaluation setup and results).
Figure~\ref{fig:taobao_unfair0_5} shows that low-end devices experience disproportionate accuracy loss when there is high system-induced data heterogeneity, similar to what was observed in the real world (Figure~\ref{fig:prod_unfair}). The same level of fairness issue cannot be observed with low system-induced data heterogeneity (Figure~\ref{fig:taobao_unfair5000}).

\subsection{Supporting Popular Tier-Aware Optimizations in FL}
\label{sec:opt}

\sys implements several commonly-used tier-aware optimizations discussed in Section~\ref{sec:bg:system_hetero}. Specifically, the current version implements (1) excluding low-end devices (\textbf{Exclude Lo}), (2) overselection and drop (\textbf{Overselect}), (3) tier-aware gradient pruning (\textbf{Prune}), (4) tier-aware gradient compression (\textbf{Quant}), and (5) tier-aware channel width reduction (\textbf{Channel}).
For simulation, a performance model for each tier was obtained from~\cite{flash}, which models the performance as a Gaussian distribution using real-world measurements.
For pruning (\textbf{Prune}), we explored random pruning that is widely used in FL~\cite{fl_comm} and do not consider Top-K pruning~\cite{topk} due to privacy concerns (Section~\ref{sec:bg:system_hetero}).
For quantization (\textbf{Quant}), we studied stochastic rounding~\cite{fl_comm, qsgd}. Given $n$ bits, stochastic rounding uniformly splits the value range between the minimum and the maximum with $2^n$ uniformly separated points $p_0$, ... $p_{2^n-1}$. If $p_k$ < $x$ < $p_{k+1}$, then we round $x$ into $p_k$ with a probability of $\frac{x - p_k}{p_{k+1} - p_k}$ and $p_{k+1}$ otherwise.
We also explore a variant that uses 1 bit to encode the sign and $n-1$ bits to encode the absolute value (\textbf{QuantS}). The variant gives us a better representation of zero, which we will show to have fairness improvement in Section~\ref{sec:eval}.
To reduce computation, we focus on varying the channel dimensions (\textbf{Channel}) without additionally using knowledge distillation~\cite{heterofl, fjord}. Knowledge distillation-based approaches~\cite{feddf, fjord, fedgkt, cronus, fedmd, faug, yaejee} require a representative public dataset which we do not assume.
We reduced channels for all but the first hidden layer, as reducing channels for the first hidden layer effectively ignores some input features.
\sys can be extended to support more optimizations.

\subsection{Quantifying Fairness}
\label{sec:bench_fairness}

To study whether an optimization strategy impacts each tier equally, we use the \emph{relative accuracy change}~\cite{hooker1, hooker2} for each tier before and after applying an optimization.
Mathematically, if model accuracy is $\beta^{t}_{p}$ for tier $t \in \{low, mid, high\}$ and an optimization $p$ ($p=0$ is no-optimization) is applied, the relative accuracy change for tier $t$ is defined as:

\begin{equation}
    \frac{\beta^{t}_{p} - \beta^{t}_{0}}{\beta^{t}_{0}}
\end{equation}

To quantify the fairness impact of an optimization $p$ across tiers, we report the \emph{maximum difference in the accuracy change (\textbf{MDAC})} between tiers. MDAC is higher if an optimization is more unfair, and 0 if perfectly fair. It is defined as:

\begin{equation}
    max(|\frac{\beta^{t_i}_{p} - \beta^{t_i}_{0}}{\beta^{t_i}_{0}} - \frac{\beta^{t_j}_{p} - \beta^{t_j}_{0}}{\beta^{t_j}_{0}}|),   t_i, t_j \in \{low, mid, high\}
\end{equation}


%% file: methodology.tex
\section{Evaluation Methodology}
\label{sec:methodology}

\subsection{Deep Learning Recommendation Models and Datasets}
\label{sec:setup:dataset-models}

\paragraph{Datasets} We study two commonly-used open-source recommendation datasets, Taobao Ad Display/Click Data~\cite{taobao} (i.e., Taobao dataset) and MovieLens-20M~\cite{movielens} (i.e., MovieLens dataset).
The Taobao dataset shows 26 million interactions (click/non-click) between 1.14 million users and 847 thousand item ads across an 8-day period. Each user has 9 sparse features (e.g., gender or occupation), each ad has one dense (price) and 5 sparse (e.g., category or brand) features, and each event has one sparse feature that encodes the "scenario"~\cite{taobao}.
The MovieLens dataset provides 20 million movie ratings for 27 thousand movies from 138 thousand users, along with the genre information for each movie. To convert it into a click/non-click dataset, we considered a 5-star rating as click and others as non-click~\cite{din}.
Following prior work~\cite{din, dien}, we did not use user ID as a user feature for privacy. Instead, we augmented the user features with the user history of previously clicked items (ads, categories, and brands for Taobao, movies for MovieLens).
For Taobao, we additionally used the day of the week information~\cite{din}.
We applied logarithm to Taobao's item price feature because the range of the value is very large, from 0.01 to 100 million.

\paragraph{Models} We evaluated two state-of-the-art deep recommender models, DLRM~\cite{dlrm,dlrm-mlperf} and DIN~\cite{din}. We did not study models that directly use user IDs, e.g., NeuMF~\cite{neumf}, for enhanced privacy.
%
DLRM~\cite{dlrm} is a model developed by Meta.
In DLRM, dense features go through a bottom MLP and are mixed with the output of the embedding tables through a pairwise dot product. The output goes through a top MLP to produce the final prediction.
DIN~\cite{din} is a model proposed by Alibaba.
In DIN, the user history features go through an \emph{attention layer} after the embedding tables, which predicts the importance of each history and gives a larger weight to the history that is more relevant to the current item. After being re-weighted, the features are concatenated with the dense features and go through an MLP for prediction.
For both models, we used the top MLP with a single hidden layer of size 256, and embedding tables with a dimension of 16. For DLRM, we used a bottom MLP with a hidden layer size 16. For DIN's attention layer, we used two hidden layers of sizes 64 and 16 and used Dice activation~\cite{din} without batch normalization. For clients, we used full-batch SGD with lr=1.0 for both datasets. For the server, we used AdaGrad with lr=0.01 for Taobao and lr=0.1 for MovieLens.

We used ROC-AUC~\cite{auc}, or AUC for short, as the accuracy metric. AUC measures the model quality well when the labels are extremely biased (e.g., when most of the ads are not clicked)~\cite{din, dien, dhe, dlrm}.
As a reference, the achieved test AUC after 1 epoch of non-FL training was 0.6096/0.6049 for the Taobao dataset with DLRM/DIN and 0.7995/0.7666 for the MovieLens dataset with DLRM/DIN, being similar to prior work~\cite{din}.
The achieved test AUC after FL training with all clients exactly once was 0.5966/0.5941 (Taobao, DLRM/DIN) and 0.7954/0.7538 (MovieLens, DLRM/DIN), which are the values used as a baseline AUC for our fairness metric (MDAC) calculation. It is hard to compare our FL results with prior work directly because no prior work trained the exact same datasets and models in an FL setup; however, the achieved AUC falls into a similar range as prior work that used similar datasets~\cite{alibaba_fl, din}.



\subsection{Tier-Aware Optimizations}

We explored six classes of tier-aware optimization techniques explained in Section~\ref{sec:opt} ({Exclude Lo}, {Overselect}, {Prune}, {Quant}, {QuantS}, {Channel}). For Prune, Quant, QuantS, and Channel, we explore three different configurations each, which impose roughly 1:2:4, 1:2:8, or 1:4:16 communication/computation overheads to low-, mid-, and high-end devices. Table~\ref{tbl:config} summarizes the 14 configurations we studied.

\begin{table}[h]
    \centering
    \begin{tabular}{| c | c | c |} \hline
        name & overhead ratio & optimization for each tier\\\hline
        Exclude Lo & & \makecell[l]{Prohibits low-end devices from FL training.}\\ \hline
        Overselect & & \makecell[l]{Selects and drops 20\% extra clients.} \\ \hline
        \multirow{3}{*}{Prune} & 1:2:4 & \makecell[l]{Prunes 75\% (low), 50\% (mid), and 0\% (high) of the gradients.} \\
         & 1:2:8 & \makecell[l]{Prunes 87.5\% (low), 75\% (mid), and 0\% (high) of the gradients.} \\
         & 1:4:16 & \makecell[l]{Prunes 93.75\% (low), 75\% (mid), and 0\% (high) of the gradients.} \\ \hline
         
         \multirow{3}{*}{\makecell[c]{Quant\\QuantS}} & 1:2:4 & \makecell[l]{Quantize the gradients using 8 (low), 16 (mid), and 32bits (high).} \\
         & 1:2:8 & \makecell[l]{Quantize the gradients using 4 (low), 8 (mid), and 32bits (high).} \\
         & 1:4:16 & \makecell[l]{Quantize the gradients using 2 (low), 4 (mid), and 32bits (high).} \\ \hline

         \multirow{3}{*}{\makecell[c]{Channel}} & 1:2:4 & \makecell[l]{Use 25\% (low), 50\% (mid), and 100\% (high) of the original channel size.} \\
         & 1:2:8 & \makecell[l]{Use 12.5\% (low), 25\% (mid), and 100\% (high) of the original channel size.} \\
         & 1:4:16 & \makecell[l]{Use 6.25\% (low), 25\% (mid), and 100\% (high) of the original channel size.} \\ \hline
    \end{tabular}
    \caption{Optimizations tested with \sys.}
    \label{tbl:config}
\end{table}
\vspace{-30pt}

\subsection{System-Induced Data Heterogeneity}
We evaluated the effect of varying levels of system-induced data heterogeneity by evaluating all the configurations on (1) random tier mapping (\textbf{Random}, no system-induced data heterogeneity), and (2) Dirichlet-based tier mapping using five different $\alpha$: \textbf{Hetero-vlow} ($\alpha=5000$), \textbf{Hetero-low} ($\alpha=5$), \textbf{Hetero-mid} ($\alpha=0.5$), \textbf{Hetero-high} ($\alpha=0.05$), and \textbf{Hetero-vhigh} ($\alpha=0.005$). The configurations represent very low to very high system-induced data heterogeneity.

%% file: eval.tex
\section{Evaluation Results}
\label{sec:eval}

Our evaluation aims to answer the following questions in the presence of realistic system-induced data heterogeneity:
\begin{itemize}
    \item How do tier-aware optimization strategies from prior literature affect fairness?
    \item How does the degree of system-induced data heterogeneity affect fairness?
    \item How do different models and datasets affect fairness?
    \item Is the best-performing optimization in terms of prediction accuracy also the best in terms of fairness?
\end{itemize}


\subsection{Fairness Impacts of Different Optimizations Under System-Induced Data Heterogeneity}

\begin{figure}
     \centering
     \begin{subfigure}[b]{0.99\textwidth}
         \centering
         \includegraphics[width=\textwidth]{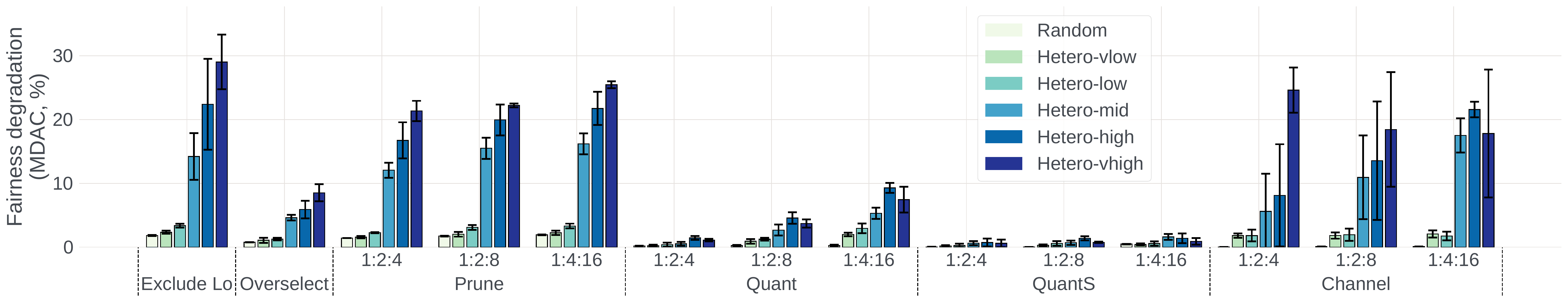}
         \caption{Taobao dataset w/ DLRM}
         \label{fig:eval1_0}
     \end{subfigure}
     \begin{subfigure}[b]{0.99\textwidth}
         \centering
         \includegraphics[width=\textwidth]{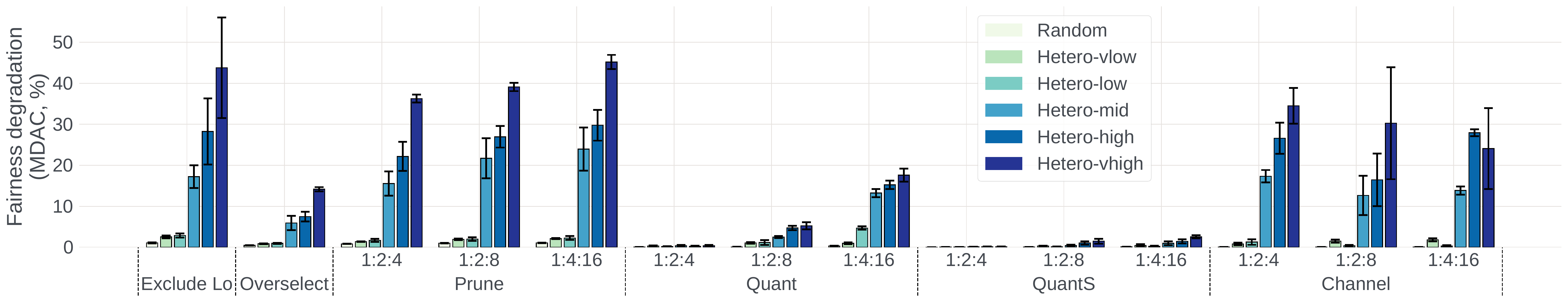}
         \caption{Taobao dataset w/ DIN}
         \label{fig:eval1_1}
     \end{subfigure}
        \caption{Different optimizations and different system-induced data heterogeneity have different fairness impacts. The figure plots the fairness implications of 14 different optimizations on 6 different heterogeneity levels on the Taobao dataset.}
        \label{fig:eval1}
\end{figure}

\begin{figure}
     \centering
     \begin{subfigure}[b]{0.98\textwidth}
         \centering
         \includegraphics[width=\textwidth]{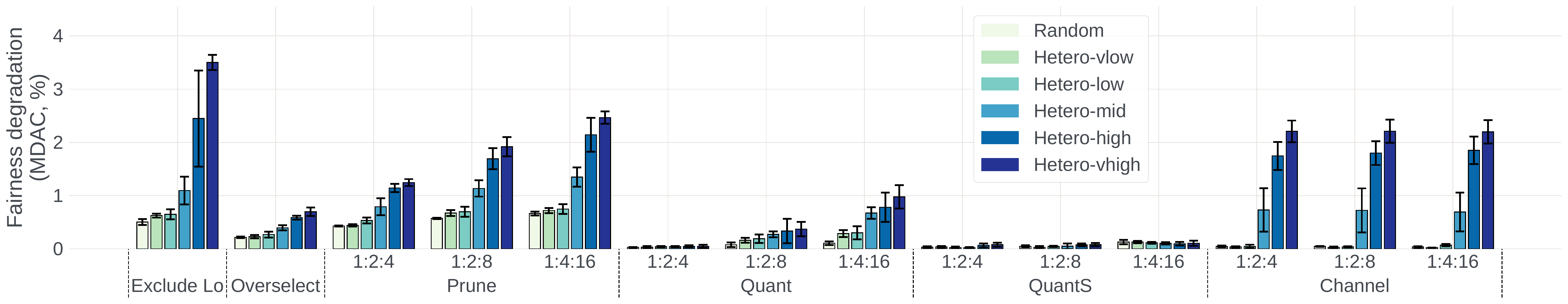}
         \caption{MovieLens-20 dataset w/ DLRM}
         \label{fig:eval2_0}
     \end{subfigure}
     \begin{subfigure}[b]{0.98\textwidth}
         \centering
         \includegraphics[width=\textwidth]{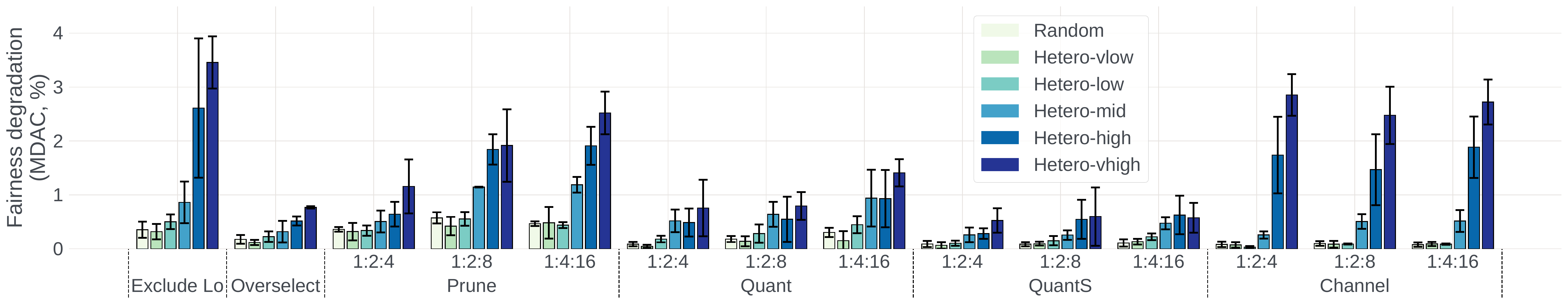}
         \caption{MovieLens-20 dataset w/ DIN}
         \label{fig:eval2_1}
     \end{subfigure}
        \caption{Different optimizations and different system-induced data heterogeneity have different fairness impacts. The figure plots the fairness implications of 14 different optimizations on 6 different heterogeneity levels on the MovieLens dataset.}
        \label{fig:eval2}
\end{figure}

Figure~\ref{fig:eval1}--\ref{fig:eval2} shows the results of training each model and dataset under the 14 optimization configurations and the 6 different system-induced data heterogeneity settings. The y-axis shows the fairness degradation (MDAC, defined in Section~\ref{sec:bench_fairness}). A larger MDAC means the optimization strategy is more unfair.

\paragraph{Takeaway 1: Optimizations cause fairness degradation}

In the presence of system-induced data heterogeneity (e.g., Hetero-vhigh/high), tier-aware optimizations may introduce significant fairness degradation.
For example, Exclude Lo, which is an optimization used by Google~\cite{gboard_hard}, caused 29--44\% MDAC with DLRM/DIN and Taobao dataset (Figure~\ref{fig:eval1}). The result means that low-end devices can suffer 29--44\% more accuracy degradation than high-end devices in the presence of high system-induced data heterogeneity.
Figure~\ref{fig:eval1}--\ref{fig:eval2} also shows that more skewed tier-aware optimizations generally lead to a higher fairness degradation. For each optimization, the bar heights generally increase as we move from a less skewed optimization (1:2:4) to a more skewed optimization (1:4:16) in each optimization group.

\paragraph{Takeaway 2: Fairness impacts change depending on the degree of system-induced data heterogeneity}

Figure~\ref{fig:eval1}--\ref{fig:eval2} shows that compared to no or very low heterogeneity cases (Random/Hetero-vlow), more fairness is hampered when the degree of system-induced data heterogeneity is higher (Hetero-vhigh/high).
Exclude Lo, for example, see more than 15.8$\times$ fairness degradation for DLRM/Taobao (MDAC 1.84\% vs. 29\%, Figure~\ref{fig:eval1_0}), and 41$\times$ for DIN/Taobao (MDAC 1.06\% vs. 43.7\%, Figure~\ref{fig:eval1_1}).
The results clearly imply that when studying tier-aware optimizations for FL, simulating realistic system-induced data heterogeneity is crucial; otherwise, one might downplay the fairness implication of an optimization by as much as 41$\times$.

\paragraph{Takeaway 3: Different optimizations have different fairness impacts}
Figure~\ref{fig:eval1}--\ref{fig:eval2} also shows that some optimizations are fairer than the others in the presence of system-induced data heterogeneity.
Take a look at Figure~\ref{fig:eval1_0}, for example. By only looking at random mapping (Random), it may seem like Channel 1:2:4 brings similar fairness concerns with QuantS 1:2:8 (MDAC 0.045\% vs. 0.044\%).
However, in the presence of system-induced data heterogeneity, QuantS 1:2:8 is much more fair than Channel 1:2:4 (MDAC 0.78\% versus 24.62\% for Hetero-vhigh, 1.41\% versus 8.13\% for Hetero-high, 0.73\% versus 5.63\% for Hetero-mid).
This result again warns that only looking at random or low system-induced data heterogeneity cases might send a misguided message when assessing the fairness of different optimizations.
Among the methods we studied, we saw Exclude Lo had the most unfair impact, while Quant/QuantS was the fairest.

\paragraph{Takeaway 4: Fairness impacts depend on the dataset/model architecture}
Comparing Figure~\ref{fig:eval1} and Figure~\ref{fig:eval2}, we can see that fairness also depends significantly on the characteristics of the dataset itself. The fairness impact of the optimizations is an order of magnitude larger for Taobao, compared to MovieLens (average MDAC 6.67\% vs. 0.64\% for DLRM + Hetero-vhigh, 6.45\% vs. 0.55\% for DLRM + Hetero-high). 
One hypothesis is that the rating of a movie is universal and easier to predict (i.e., a good movie is considered good by everybody) compared to Ads-clicks and, therefore, can be learned better even under a high degree of system-induced data heterogeneity.
Similarly, comparing Figure~\ref{fig:eval1} and Figure~\ref{fig:eval2} reveals that DIN experiences slightly higher fairness degradation compared to DLRM (e.g., average MDAC 6.67\% vs. 8.96\% for DLRM + Hetero-vhigh). We can conclude that the fairness impact of different tier-aware optimizations heavily depends on both datasets and model architectures.

%
%

\paragraph{Takeaway 5: Quantization with separate sign encoding improves fairness}
Comparing Quant with QuantS shows that QuantS impacts fairness much less. When comparing across all the configurations for high system-induced data heterogeneity scenarios (Hetero-vhigh/high/mid), QuantS 1:2:4 improves the fairness by 1.4 -- 1.7$\times$ compared to Quant 1:2:4, QuantS 1:2:8 by 2.8--4$\times$ compared to Quant 1:2:8, and QuantS 1:4:16 by 4.9--5.9$\times$ compared to Quant 1:4:16.
The reason is that while optimizations like pruning only lose gradient information within the tier if applied to a certain tier, quantization actually introduces \emph{noise} in the gradient that can affect the model quality of other tiers. Particularly for embedding tables, gradients for the table entries that were not accessed by a certain client must be close to zero for that client. However, quantization may make these gradients non-zero as it will round zero into a nearby quantized value, introducing noise to un-accessed embedding entries.
Because quantization with a separate sign encoding can better encode zero, it shows significantly better fairness results.
The finding demonstrates a scenario where researchers can evaluate the fairness implications of their optimization proposals and modify their optimizations using the proposed \sys framework.

\subsection{Zooming into Each Optimization's Impact on Each Tiers}

\begin{figure}
    \centering
    \includegraphics[width=\textwidth]{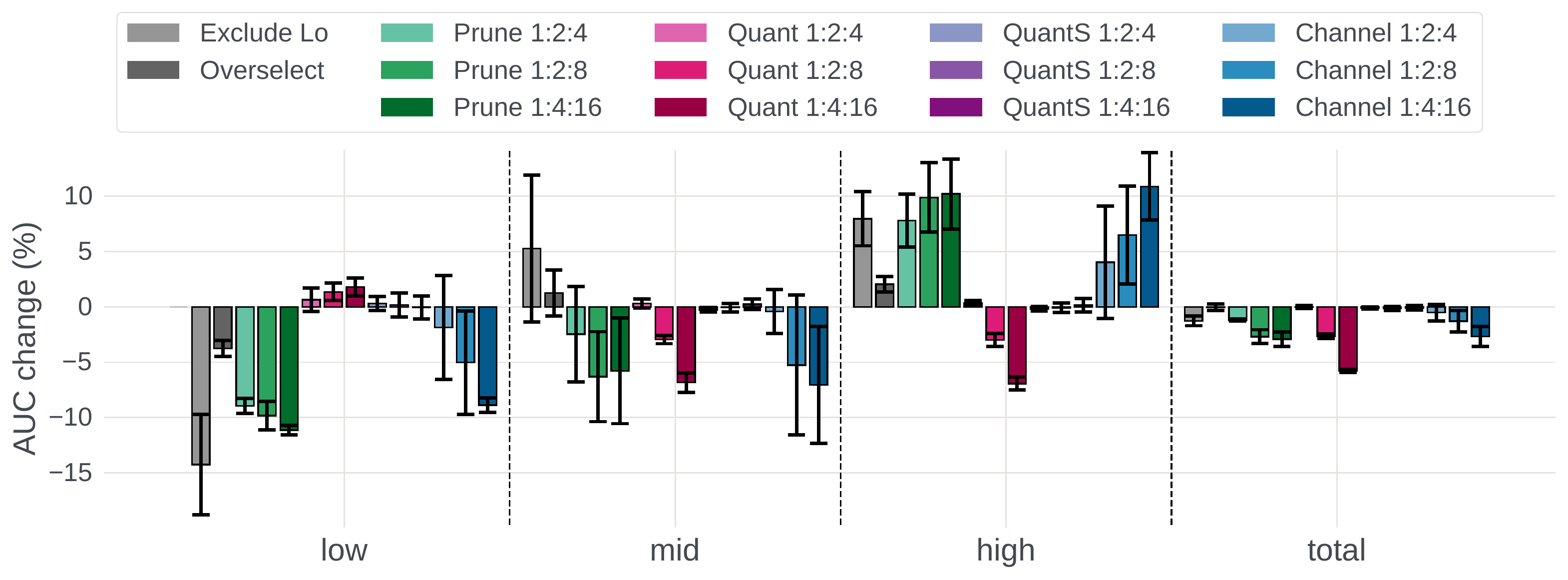}
    \caption{Different tiers are impacted differently for each optimization. The figure shows the case of training the Taobao dataset with DLRM with Hetero-high ($alpha=0.05$) in more detail, plotting the AUC change for each tier separately. The plot shows that while most of the studied optimizations degrade low-end devices' AUC more severely, quantization degrades mid/high-end devices' AUC more due to low-end devices' quantized gradient corrupting the updates from other tiers.}
    \label{fig:eval3}
\end{figure}

Figure~\ref{fig:eval3} illustrates the AUC change for each tier separately for one representative configuration --- training DLRM with the Taobao dataset under Hetero-high. The results for the other setups showed similar trends and were omitted for space reasons. Zooming into the effect on each tier separately highlights additional interesting observations.

\paragraph{Takeaway 6: Quantization benefits low-end devices, while all other optimizations punish low-end devices.}
As expected, most of the tier-aware optimization strategies degrade the model accuracy of the low-end devices disproportionately, because optimizations are more aggressively applied to resource-constrained, low-end devices.
However, quantization degrades the model accuracy of mid/high-end devices more.
The reason for the unexpected suffering of mid/high-end devices using quantization is again because quantized gradients of the low-end devices pollute the model updates of mid/high-end devices, especially from the embedding tables.

\paragraph{Takeaway 7: The best-accuracy optimization is not always the best-fairness optimization}
When comparing the overall AUC degradation (the \textbf{total} bar group in Figure~\ref{fig:eval3}) with the fairness impact of each optimization (Figure~\ref{fig:eval1}), we can see that the optimizations that lead to minimal overall AUC degradation do not always coincide with optimizations that are the fairest.
For example, Exclude Lo, which is one of the most unfair optimizations, shows reasonable AUC degradation (-1.26\%) that is better than Quant 1:2:8 (-2.64\%) and Quant 1:4:16 (-5.79\%). However, Quant 1:2:8 and Quant 1:4:16 are much fairer (MDAC 4.59\% and 9.3\%, Figure~\ref{fig:eval1_0}) than Exclude Lo (MDAC 22.4\%, Figure~\ref{fig:eval1_0}).
This result indicates that only evaluating the overall accuracy after applying an optimization, as in the prior work~\cite{heterofl, fjord}, may present an incomplete picture. Both the final model accuracy and per-tier fairness (MDAC) must be considered to understand the overall design and optimization space better.



%% file: related.tex
\section{Additional Related Work}

\paragraph{Fairness in ML}

Remotely related, many studies showed that applying optimizations on a trained model can disproportionately harm minorities in the dataset~\cite{hooker1, hooker2}.
A recent public study also showed that using smartphone data to train ML models can produce a model unfair towards groups without smartphones~\cite{boston_street_bump}.
Our work shows how applying tier-aware optimizations during FL can impact groups with low-tier devices, studying distinguished aspects from these studies.
Whether prior debiasing solutions~\cite{fairness0, fairness1, fairness2} can be applied to our setup is an interesting future work.


\paragraph{FL simulation frameworks} Several simulation frameworks exist for FL~\cite{leaf, flash, fedscale, flower, flsim, tff, pysyft, ibmfl, fedjax}. Unlike \sys, none of the prior simulators that we are aware of support simulating system-induced data heterogeneity, even the frameworks that focus on realistic system-heterogeneity simulation~\cite{fedscale, flash}.
%
%
These other simulators can adopt the core idea of \sys's system-induced data heterogeneity simulation and implement it in their framework.

\paragraph{Other FL optimizations for system heterogeneity}
In addition to the tier-aware optimizations we discuss in Section~\ref{sec:bg:system_hetero}, other work proposed complementary solutions to tackle the system heterogeneity problem in FL.
AutoFL~\cite{autofl} and OORT~\cite{oort} use ML-based client selection taking into account clients' system heterogeneity.
FedBuff~\cite{fedbuff} and Papaya~\cite{papaya} implement asynchronous FL to mitigate the stragglers' effect. FedBuff still down-weighs slower clients' updates~\cite{fedbuff}.
These systems were evaluated assuming no data-system inter-dependence, and studying the effect of system-induced data heterogeneity for these proposals will be interesting future work.

\paragraph{Memory-efficient recommender systems}

Training recommender systems on-device requires the models to be memory-efficient. Reducing MLP layers can be done by reducing the channel dimension~\cite{heterofl, fjord, expanding_reach}, which was studied in this paper. Additionally, a set of techniques were proposed to reduce memory usage of embedding tables on the server-side~\cite{qr, tt, ttrec, dhe}. Recent work~\cite{fedrecon} also proposed reconstructing embedding layers on-device during FL to reduce the memory footprint of a large table.
These techniques have not been evaluated in the presence of system-induced data heterogeneity in FL.

%% file: conclusion.tex
\section{Conclusion}

To enhance data privacy in recommender systems, federated learning has emerged as an effective mechanism. Despite a plethora of prior works on FL, an important characteristic of the real-world environment has not yet been considered. In this work, we shed light on the under-explored aspect of the inter-dependence between system and data heterogeneity --- that has been considered individually but not in conjunction by most (if not all) prior work in the FL space. 
Based on the statistical observations from the real-world environment, we design a new statistical framework to model and evaluate the impact of system-induced data heterogeneity for federated recommendation learning. Our evaluation results demonstrate that fairness can be severely affected under realistic system-induced data heterogeneity, and modeling the inter-dependence is essential to understanding the true fairness impact of model optimization strategies.

%% file: acks.tex
We would like to thank Kamalika Chaudhuri and Edward Suh for the invaluable discussion regarding the paper's direction. We thank Ilias Leontiadis, Shripad Gade, Mani Malek, Fangzhou Xu, Vlad Grytsun, and Shuaiwen Wang for their help in conducting the experiments with the production data. We also thank Ashkan Yousefpour, Sayan Ghosh, Hongyuan Zhan, Kaikai Wang, Dzmitry Huba, and Meisam Hejazi nia, who helped us understand the operation of open-source and production federated learning system. We thank Pegah T. Afshar, Milan Shen, and Kim Hazelwood for supporting the work.